\documentclass[aps,showpacs,twocolumn,superscriptaddress]{revtex4-1}

\usepackage{dcolumn}
\usepackage{bm}
\usepackage{color}
\usepackage{array}
\usepackage{mathrsfs}
\usepackage{ulem} 
\usepackage{amsmath}
\usepackage{amssymb}
\usepackage{url}
\usepackage{epsfig}

\begin{document}


\title{Ground state phase diagram of Kitaev-Heisenberg model \\ on honeycomb-triangular lattice}


\author{Masanori Kishimoto}
\email[e-mail:]{1517611@ed.tus.ac.jp}
\affiliation{Department of Applied Physics, Tokyo University of Science, Tokyo 125-8585, Japan}

\author{Katsuhiro Morita}
\email[e-mail:]{katsuhiro.morita@rs.tus.ac.jp}
\affiliation{Department of Applied Physics, Tokyo University of Science, Tokyo 125-8585, Japan}

\author{Yukihiro Matsubayashi}
\affiliation{Department of Applied Physics, Tokyo University of Science, Tokyo 125-8585, Japan}

\author{Shigetoshi Sota}
\affiliation{ Computational  Materials  Science  Research  Team, RIKEN Center for Computational Science (R-CCS),  Kobe,  Hyogo  650-0047,  Japan}

\author{Seiji Yunoki}
\affiliation{ Computational  Materials  Science  Research  Team, RIKEN Center for Computational Science (R-CCS),  Kobe,  Hyogo  650-0047,  Japan}
\affiliation{Computational Condensed Matter Physics Laboratory, RIKEN, Wako, Saitama 351-0198, Japan}
\affiliation{Computational Quantum Matter Research Team, RIKEN Center for Emergent Matter Science (CEMS), Wako, Saitama 351-0198, Japan}

\author{Takami Tohyama}
\affiliation{Department of Applied Physics, Tokyo University of Science, Tokyo 125-8585, Japan}


\date{\today}

\begin{abstract}
The Kitaev-Heisenberg model defined on both honeycomb and triangular lattices has been studied intensively in recent years as a possible model to describe spin-orbital physics in iridium oxides. In the model, there are many phases characteristic for each lattice. However, there is no study how the phases in the two lattices merge each other when geometry changes from honeycomb lattice to triangular lattice. We investigate the ground state of the Kitaev-Heisenberg model defined on the system connecting the honeycomb and triangular lattices, named a honeycomb-triangular lattice. We obtain a ground state phase diagram of this model with classical spins by using the Luttinger-Tisza method and classical Monte Carlo simulation. In addition to known phases in the honeycomb and triangular lattices, we find coexisting phases consisting of the known phases. Based on the exact diagonalization and density-matrix renormalization group calculations for the model with quantum spins, we find that quantum fluctuations less affect on the phase diagram.   
\end{abstract}

\maketitle

\section{Introduction}
\label{Sec1}

Strong spin-orbital interaction plays an important role in iridium oxides by entangling spin and orbital degrees of freedom in $5d$ electrons. 
As a result, the iridates can be effectively described by Kitaev-type interactions together with isotropic Heisenberg-type interactions~\cite{jackeli2009}.
When the Kitaev interactions, which have been originally studied by Kugel and Khomskii on the basis of the compass model~\cite{kugel1982, khaliullin2005, nussinov2015}, are dominating on a honeycomb lattice, there appears a Kitaev spin liquid~\cite{kitaev2006}.
This fact has brought intensive studies of the Kitaev--Heisenberg (KH) model on the honeycomb lattice, which has been expected to be an effective model for (Na,Li)$_2$IrO$_3$~\cite{chaloupka2010, jiang2011, reuther2011, okamoto2013, schaffer2013, chaloupka2013, price2013, sela2014}. 
To describe (Na,Li)$_2$IrO$_3$ more realistically, extended versions of the KH model with further-neighbor interactions~\cite{kimchi2011, singh2012, choi2012, reuther2014}, and additional anisotropic interactions~\cite{bhattacharjee2012, katukuri2014, rau2014, rau2014b, yamaji2014, sizyuk2014, kimchi2015, shinjo2015, chaloupka2015, okubo2017} have been studied. 

In addition to the honeycomb lattice, the KH model on a triangular lattice has attracted attention from a theoretical viewpoint~\cite{kimchi2014,becker2015,rousochatzakis2016,shinjo2016}, since it has both geometrical frustration and Kitaev-type frustration that breaks the SU(2) spin symmetry. 
In addition, from the experimental side, Ba$_3$IrTi$_2$O$_9$ has been suggested as a possible spin-liquid material with a frustrated triangular-lattice structure containing compasslike magnetic interactions~\cite{dey2012,lee2017} and been discussed in connection to the KH model~\cite{catuneanu2015}.

Though the KH model cannot properly describe existing iridates, the model has attracted a lot of attention in terms of  phase diagrams on the honeycomb lattice~\cite{chaloupka2010,jiang2011,chaloupka2013} and the triangular lattice~\cite{kimchi2014,becker2015,rousochatzakis2016,shinjo2016}.
In the honeycomb lattice, there are the N\'eel, spin liquid, zigzag, ferromagnetic (FM), and stripy phases, while in the triangular lattice there are the 120-degree ordered, $\mathbb {Z}_2$ vortex crystal ($\mathbb {Z}_2$VC), nematic, dual $\mathbb {Z}_2$VC, and dual FM phases.
These phases appearing in the two lattices are not the same.
Therefore, it is interesting to know how the phases evolve and merge each other when geometry changes from honeycomb lattice to triangular lattice. 

In this paper, we construct a model connecting the honeycomb and triangular KH lattices and examine the ground state of the classical system by the Luttinger-Tisza (LT) method~\cite{luttinger1946,litvin1974} and the classical Monte Carlo (MC) simulation.
We find coexisting phases in the honeycomb-triangular lattice, which are composed of the known phases. 
The phase boundaries in the classical phases survive even though quantum fluctuations are introduced as evidenced from the exact diagonalization (ED) and density-matrix renormalization group (DMRG) calculations for small systems. 
 
This paper is organized as follows. The KH model on the honeycomb-triangular lattice is introduced in Sec.~\ref{Sec2}. In Sec.~\ref{Sec3}, the methods used in this paper, i.e., the LT method, classical MC simulation, ED, and DMRG are described. The classical ground state phase diagram is shown in Sec.~\ref{Sec4} with the emphases of the nature of coexisting phases. In Sec.~\ref{Sec5}, we examine the ground state of the same model for quantum spins by the ED and DMRG methods and show a quantum phase diagram. Finally, a summary is given in Sec.~\ref{Sec6}.

\section{Model}
\label{Sec2}

\begin{figure}[tb]
  \begin{center}
	\epsfig{file=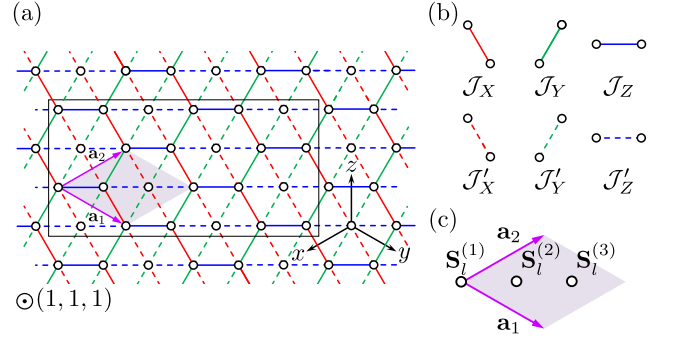, width=86mm}
    \caption{(a) Honeycomb-triangular lattice on the plane perpendicular to $(x,y,z)=(1,1,1)$, where $(x,y,z)$ is the orthogonal coordinate system used for the spin space. The red, green, and blue lines are perpendicular to the $x$, $y$, and $z$ axes, respectively. The solid lines form the honeycomb lattice, while the dashed lines connect the honeycomb sites with a central site. The shaded area represents a unit cell. The rectangle denoted by black solid line presents a 24-site lattice with 8 unit cells used in ED calculations.  (b) Six types of nearest-neighbor interactions appearing in (a). The red solid, green solid, blue solid, red dashed, green dashed, and blue dashed lines represent the matrix of nearest-neighbor interactions, $\mathcal{J}_X$, $\mathcal{J}_Y$, $\mathcal{J}_Z$, $ \mathcal{J}' _X$, $ \mathcal{J}' _Y$, and $ \mathcal{J}' _Z$, respectively. The matrices of the interactions contain only diagonal elements (see the main text). (c) The definition of unit cell. ${\bf a}_1$ and ${\bf a}_2$ are two primitive translation vectors. $\mathbf{S}_l^{(m)}$ represents a spin at the sublattice $m$ in the unit cell $l$.}
    \label{fig:model}
  \end{center}
\end{figure}

The Hamiltonian of the Kitaev-Heisenberg model on the honeycomb-triangular lattice is given by
\begin{equation}
  \mathcal{H}= \sum _{\langle i,j \rangle} {\bf S}_i ^{\rm T}  \mathcal{J}_{i,j}{\bf S}_j \label{eq:hamiltonian_1},
\end{equation}
where ${\bf S}_i$ is a classical spin ${\bf S}_i=(S_i^x\  S_i^y\ S_i^z)^{\rm T} \in \mathbb{R} ^3$ with $|{\bf S}_i|=1$ (a spin operator with $S=1/2$) at site $i$ for classical (quantum) system. $ \mathcal{J}_{i,j}$ represents the nearest-neighbor interaction given by the matrix form: for the bond connecting two honeycomb sites, $\mathcal{J}_X$, $\mathcal{J}_Y$, and $\mathcal{J}_Z$ are perpendicular to the $x$, $y$, and $z$ directions in the spin space, respectively, while for the bond connecting a honeycomb site to a neighboring central site, $ \mathcal{J}' _X$, $ \mathcal{J}' _Y$, and $ \mathcal{J}' _Z$, are assigned as shown in Figs.~\ref{fig:model}(a) and \ref{fig:model}(b). All of the matrices have only diagonal elements with $\mathcal{J}_X=\mathrm{diag}(J+K, J, J)$, $\mathcal{J}_Y = \mathrm{diag}(J, J+K, J)$, $\mathcal{J}_Z=\mathrm{diag}(J, J, J+K)$, $ \mathcal{J}' _X=\alpha \mathcal{J}_X$, $ \mathcal{J}' _Y=\alpha \mathcal{J}_Y$, and $ \mathcal{J}' _Z= \alpha \mathcal{J}_Z$ ($0 < \alpha \le 1$), where $K$ and $J$ correspond to the coefficient  of the Kitaev and Heisenberg terms, respectively. The parameter $\alpha$ determines how a given lattice is close to triangular lattice, i.e., $\alpha = 1$ corresponds to a triangular lattice, while $\alpha \to 0^+$ corresponds to a honeycomb lattice. We note that there is the Klein duality~\cite{kimchi2014} in this model, which transforms $(J,K) \mapsto ( \tilde{J}, \tilde{K})=(-J, 2J+K)$. 

\section{Method}
\label{Sec3}

\subsection{Luttinger-Tisza (LT) method}
In order to analyze the classical ground state, we introduce a unit cell as shown in Fig.~\ref{fig:model}(c) and the Fourier transform of spins defined as 
\begin{equation}
  S_l ^{(m) \gamma} \equiv \frac{1}{\sqrt{N}} \sum _{\bf q} e^{i {\bf q} \cdot {\bf r}_l ^{(m)}} S_{\bf q} ^{(m) \gamma}, \label{eq:fourier}
\end{equation}
where $S_l ^{(m) \gamma}$ denotes the $\gamma$ component of the spin $\mathbf{S} _l ^{(m)}$ at ${\bf r}_l ^{(m)}$ belonging to the sublattice $m$ in the unit cell $l$, and $N$ is the number of unit cells. 
From Eqs.~(\ref{eq:hamiltonian_1}) and (\ref{eq:fourier}), the total energy of the system is given by
\begin{equation}
  E_{\rm tot}= \frac{1}{2} \sum _{\gamma \in \{ x,y,z \}} \sum _{{\bf q} \in V_{\rm BZ}} {\bf S}_{-{\bf q}} ^{\gamma {\rm T}} \mathcal{J}_{\bf q} ^{\gamma} {\bf S}_{{\bf q}} ^{\gamma}, \label{eq:Etot}
\end{equation}
where the summation $\sum _{{\bf q} \in V_{\rm BZ}}$ is taken over the wavevectors in the first Brillouin zone $V_{\rm BZ}$. 
The three-dimensional vector ${\bf S}_{{\bf q}} ^{\gamma}$ is defined as ${\bf S}_{\bf q} ^{\gamma} \equiv (S_{\bf q} ^{(1) \gamma} \ S_{\bf q} ^{(2) \gamma} \ S_{\bf q} ^{(3) \gamma})^{\rm T}$ and the Fourier transform of the interaction is given by
\begin{equation}
  \mathcal{J}_{\bf q} ^{\gamma} = \left(
    \begin{array}{ccc}
      0 & A_{\bf q} ^{\gamma} & \alpha A_{\bf q} ^{\gamma *} \\
      A_{\bf q} ^{\gamma *} & 0 & \alpha A_{\bf q} ^{\gamma} \\
      \alpha A_{\bf q} ^{\gamma } & \alpha A_{\bf q} ^{\gamma *} & 0
    \end{array}
  \right)
\end{equation}
with
\begin{eqnarray*}
  A_{\bf q} ^x &=& (J+K)e^{i \frac{2 \pi}{3}(-2m_1+m_2)} +Je^{i \frac{2 \pi}{3}(m_1-2m_2)} \nonumber \\
  & &  \ \ +Je^{i \frac{2 \pi}{3}(m_1+m_2)}, \nonumber \\
  A_{\bf q} ^y &=& Je^{i \frac{2 \pi}{3}(-2m_1+m_2)} +(J+K)e^{i \frac{2 \pi}{3}(m_1-2m_2)} \nonumber \\
  & &  \ \ +Je^{i \frac{2 \pi}{3}(m_1+m_2)}, \nonumber \\
  A_{\bf q} ^z &=& Je^{i \frac{2 \pi}{3}(-2m_1+m_2)} +Je^{i \frac{2 \pi}{3}(m_1-2m_2)} \nonumber \\
  & &  \ \ +(J+K)e^{i \frac{2 \pi}{3}(m_1+m_2)}, \label{eq:Jdef}
\end{eqnarray*}
where $m_1$ and $m_2$ are integers that specify the wavevector as ${\bf q}=m_1{\bf b}_1+m_2{\bf b}_2$ with $\mathbf{b}_1$ and $\mathbf{b}_2$ being the reciprocal lattice vectors.

To find the ground state, we need to minimize $E_{\rm tot}$ with respect to $\{ {\bf S}_{\bf q} ^{\gamma} \}$ under the local constraints:
\begin{equation}
|{\bf S} _l ^{(m)}|^2=1\ \ \ (l=1,2,\ldots,N; \ m=1,2,3).
\label{eq:local}
\end{equation}
The LT method~\cite{luttinger1946, litvin1974} is useful for such a minimization problem. 
The local constraint is often replaced by the single global constraint $\sum _{l,m}|{\bf S} _l ^{(m)}|^2=3N$. 
However, this approach generally fails when the unit cell contains non-equivalent sites like the present model (\ref{eq:hamiltonian_1}). 
Therefore, we consider minimizing $E_{\rm tot}$ under the improved global constraints~\cite{litvin1974, yoshimori1959, freiser1961}: 
\begin{equation}
  \sum _{l=1} ^{N} \left( |{\bf S}_l ^{(1)}|^2 + |{\bf S}_l ^{(2)}|^2 \right) = 2N,
 \ \ \sum _{l=1} ^{N} |{\bf S}_l ^{(3)}|^2=N. \label{eq:global_2}
\end{equation}
Assuming that ${\bf S}_{\bf q} ^{\gamma}$ has finite magnitude, we obtain the energy per unit cell
\begin{equation}
  E_{\rm unit} = \frac{1}{2} \left[ 3 \lambda - (1- \alpha ^2) \frac{|A_{\bf q} ^{\gamma}|^2}{\lambda} \right], \label{eq:Eunit_min}
\end{equation}
where $\lambda$ is obtained as the solution of the following equation: 
\begin{eqnarray}
  \lambda ^4 - (\alpha ^2 + 2)|A_{\bf q} ^{\gamma}|^2 \lambda ^2 &-&2 \alpha ^2 \Re{[A_{\bf q} ^{\gamma 3}]} \lambda \nonumber \\
  &+& (1- \alpha ^2)|A_{\bf q} ^{\gamma}|^4=0. \label{eq:lambda_4}  
\end{eqnarray}
Thus, we need to find $\{ ({\bf q}', \gamma ') \} \ ( \subset V_{\rm BZ} \times \{ x,y,z \} )$ whose elements give the same $\lambda$, minimizing $E_{\rm unit}$. 
For $({\bf q}, \gamma ) \notin \{ ({\bf q}', \gamma ') \}$, ${\bf S}_{{\bf q}} ^{\gamma}$ has no magnitude, while for $({\bf q}, \gamma ) \in \{ ({\bf q}', \gamma ') \}$, ${\bf S}_{\bf q} ^{\gamma}$ satisfies
\begin{equation}
  \mathcal{J}_{\bf q} ^{\gamma}{\bf S}_{\bf q} ^{\gamma} = \Lambda {\bf S}_{\bf q} ^{\gamma} \label{eq:lambda_eq}
\end{equation}
with
\begin{equation}
  \Lambda \equiv {\rm diag}( \lambda , \lambda , 2(E_{\rm unit} - \lambda )). 
\end{equation}
When this solution satisfies the local constraints (\ref{eq:local}), it is the ground state.

We numerically solve Eq.~(\ref{eq:lambda_4}) to find $\{({\bf q}', \gamma ')\}$.
However, since the local constraints (\ref{eq:local}) are not necessarily satisfied when ${\bf q}'$ is incommensurate to the lattice, we use the classical MC simulation for such cases.

\subsection{Classical Monte Carlo (MC) simulation}
Our classical MC simulation is based on the single update heat-bath method combined with the over-relaxation technique and the temperature-exchange method.
After performing the MC simulation with finite but low temperature $T$, we obtain the ground state by updating the state until the energy converges at $T=0$.
We use lattices with periodic boundary conditions (PBC), containing $L \times L \times 3$ sites, where the maximum number of $L$ is 72.

We calculate order parameters after obtaining the ground state. In the honeycomb lattice, there are several ordered phases. We define the corresponding order parameters in the followings.

For the N\'eel state in the honeycomb lattice (H-N\'eel),  the order parameter is defined as
\begin{equation}
  O_{\rm H-N\acute{e}el} \equiv \frac{1}{(2N)^2} \left| \sum _{l=1} ^{N} \left( {\bf S}_l ^{(1)} - {\bf S}_l ^{(2)}  \right) \right| ^2 \label{eq:order_neel}
\end{equation}
by using two spins, $\mathbf{S}_l ^{(1)}$ and $\mathbf{S}_l ^{(2)}$, on the honeycomb sites. 

For the stripy state, the order parameter reads
\begin{equation}
  O_{\rm stripy} \equiv \frac{1}{(3N)^2} \left( |M_{\rm s} ^x|^2+|M_{\rm s} ^y|^2+|M_{\rm s} ^z|^2 \right) \label{eq:order_stripy}
\end{equation}
with
\begin{eqnarray*}
  M_{\rm s} ^x &=& \sum _l \sum _{m=1} ^3 S_l^{(m)x} (-1)^{n_1+m}, \\
  M_{\rm s} ^y &=& \sum _l \sum _{m=1} ^3 S_l^{(m)y} (-1)^{n_2+m}, \\
  M_{\rm s} ^z &=& \sum _l \sum _{m=1} ^3 S_l^{(m)z} (-1)^{n_1+n_2},
\end{eqnarray*}
where $n_1$ and $n_2$ specify the position vector at $m=1$ in the unit cell $l$ through ${\bf r}_l ^{(1)}=n_1{\bf a}_1+n_2{\bf a}_2$. We note that the spin at the central site, $m=3$, inside the honeycomb ring is included in the stripy order parameter.

In the order parameter of the zigzag phase for the honeycomb lattice (H-Zigzag), we exclude the $m=3$ site and define 
\begin{equation}
  O_{\rm H-zigzag} \equiv \frac{1}{(2N)^2} \left( |M_{\rm z} ^x|^2+|M_{\rm z} ^y|^2+|M_{\rm z} ^z|^2 \right)  \label{eq:order_zigzag}
\end{equation}
with
\begin{eqnarray*}
  M_{\rm z} ^x &=& \sum _l \sum _{m=1} ^2 S_l^{(m)x} (-1)^{n_1}, \\
  M_{\rm z} ^y &=& \sum _l \sum _{m=1} ^2 S_l^{(m)y} (-1)^{n_2}, \\
  M_{\rm z} ^z &=& \sum _l \sum _{m=1} ^2 S_l^{(m)z} (-1)^{n_1+n_2+m}. 
\end{eqnarray*}

In the triangular lattice, there is the nematic phase whose order parameter is defined as
\begin{equation}
  O_{\rm nematic} \equiv \frac{1}{(3N)^2} \left( |M_{\rm n} ^x|^2+|M_{\rm n} ^y|^2+|M_{\rm n} ^z|^2 \right)  \label{eq:order_nematic}
\end{equation}
with
\begin{equation*}
  |M_{\rm n} ^{\gamma}| = \sum_{\{\mu_\gamma\}} \left| \sum _{j\in\mu_\gamma} S_j ^{\gamma}(-1)^j \right| \ \ \ ( \gamma \in \{ x,y,z \} ), 
\end{equation*}
where $j$ runs over all sites in the chain $\mu_\gamma$ connected by the interactions $\mathcal{J}_X$ and $ \mathcal{J}' _X$ for $\gamma = x$, $\mathcal{J}_Y$ and $ \mathcal{J}' _Y$ for $\gamma = y$,  and $\mathcal{J}_Z$ and $ \mathcal{J}' _Z$ for $\gamma = z$ (see Fig.~\ref{fig:model}), and $(-1)^j$ represents alternating sign along the chain. The summation $\{\mu_\gamma\}$ runs over all possible $\mu_\gamma$ chains in the lattice.

The FM order parameter is defined as
\begin{equation}
  O_{\rm FM} \equiv \frac{1}{(3N)^2} \left| \sum _{l=1} ^{N} \left( {\bf S}_l ^{(1)} + {\bf S}_l ^{(2)}  + {\bf S}_l ^{(3)} \right) \right| ^2.
\label{eq:order_FM}
\end{equation}

\subsection{Exact diagonalization (ED) and density-matrix renormalization group (DMRG)}
In order to obtain the ground state phase diagram of the Hamiltonian (\ref{eq:hamiltonian_1}) with quantum spins (spin-1/2), we perform the Lanczos-type ED for a 24-site lattice with PBC, which contains eight unit cells as depicted in Fig.~\ref{fig:model}. The lattice is not of square shape but of rectangular shape. To compare ED results with classical ones, we calculate static spin structure factor $S(\mathbf{q})$ defined as
\begin{equation}
S(\mathbf{q})\equiv\sum_\gamma S^\gamma(\mathbf{q})=\sum_\gamma\sum_{m',m}\left<0\right| S_{-\mathbf{q}}^{(m')\gamma}S_\mathbf{q}^{(m)\gamma}\left| 0\right>,
\label{eq:Sq}
\end{equation}
where $\left| 0\right>$ represents the ground state.

\begin{figure*}[tb]
  \begin{center}
    \epsfig{file=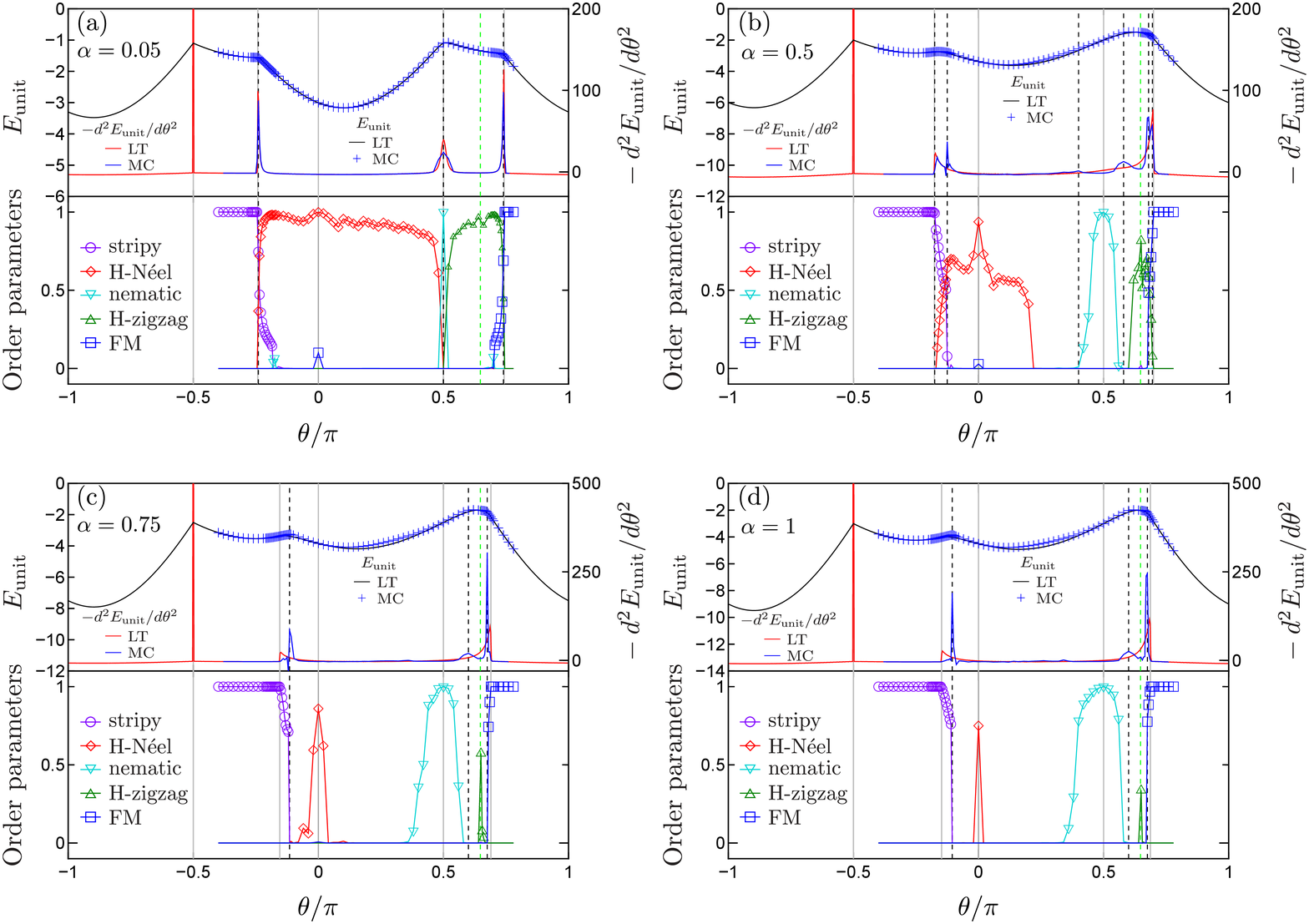, width=180mm}
    \caption{Classical ground state energy per unit cell $E_{\rm unit}$ and second derivative of $E_{\rm unit}$ with respect to $\theta$, $-d^2 E_{\rm unit}/d \theta ^2$, obtained by the LT method and the classical MC simulation for a $72 \times 72 \times 3$-site lattice with PBC, together with the order parameters ($O_\mathrm{stripy}$, $O_\mathrm{N\acute{e}el}$, $O_\mathrm{nematic}$, $O_\mathrm{H-zigzag}$, and $O_\mathrm{FM}$) obtained by the classical MC simulation. (a) $\alpha=0.05$, (b) $\alpha=0.5$, (c) $\alpha=0.75$, and (d) $\alpha=1$. The gray solid vertical lines denote the phase boundaries obtained by the LT method, while the black dashed vertical lines are obtained by the classical MC simulation. The green dashed vertical lines represent the dual point of $\theta=0$.}
    \label{fig:energy}
  \end{center}
\end{figure*}

To check the size-dependence of the ground state energy as well as $S(\mathbf{q})$, we perform DMRG calculations for a $12\times 6$-site system with open boundary condition along the 12-site direction and PBC along the 6-site direction, i.e., cylindrical boundary condition (CBC), where twenty-one hexagons are involved. In our DMRG, we construct a snakelike one-dimensional chain, and use the truncation number $m=1500$, and the resulting truncation error is less than $5\times 10^{-6}$. To compare $S(\mathbf{q})$ by DMRG with that by ED, we use $\mathbf{q}$ values determined by the standard PBC.

\section{Classical ground state phase diagram}
\label{Sec4}

We introduce angle parameter $\theta$ that determines the Heisenberg interaction $J$ and the Kitaev interaction $K$ through $J= I_0 \cos{\theta}$ and $K= I_0 \sin{\theta}$~\cite{chaloupka2013}, where $I_0$ is the unit of energy, taken to be $I_0=1$ in this paper. Changing both $\theta$ and $\alpha$, we construct the phase diagram by examining the ground state energy as well as the order parameters obtained by classical MC simulations. 

The ground state energy per unit cell, $E_\mathrm{unit}$, calculated by the LT method and classical MC simulation together with its second derivative, $-d^2E_\mathrm{unit}/d\theta^2$, are plotted in Fig.~\ref{fig:energy} as a function of $\theta$ for four values of $\alpha$.  We note that $E_\mathrm{unit}$ at $\alpha=0^+$ (not shown) and $\alpha=1$ [Fig.~\ref{fig:energy}(d)] agrees well with previously reported results for the honeycomb and triangular lattices~\cite{rau2014,rousochatzakis2016,becker2015}, respectively. As shown in Fig.~\ref{fig:energy}, the LT and MC methods give slightly different $E_\mathrm{unit}$ in the range of $0<\theta <\pi/2$. This difference mainly comes from the breakdown of the local constraints (\ref{eq:local}) in the LT method. Therefore, $E_\mathrm{unit}$ given by the classical MC simulation is more reliable. As a result, the phase boundaries giving rise to singular behaviors in $-d^2E_\mathrm{unit}/d\theta^2$ are slightly different between the two methods.

At $\alpha=0.05$ shown in Fig.~\ref{fig:energy}(a), we find four singular points in $-d^2E_\mathrm{unit}/d\theta^2$. The singularity at $\theta=-\pi/2$ corresponds to the first-order phase transition from the FM phase to the stripy phase across FM Kitaev point. At $\theta=\pi/2$, on the other hand, the singular behavior is different from that at $\theta=-\pi/2$. In fact, there is the nematic order at $\theta=\pi/2$ as shown in the lower panel. Two other singularities at $\theta\approx-0.25\pi$ and $0.75\pi$ are interrelated by the Klein duality and correspond to the first-order transition from the H-N\'eel to stripy states and from the H-Zigzag to FM states, respectively. We notice that both the H-N\'eel and stripy orders coexist near $\theta=-0.25\pi$ as shown in the lower panel. A similar coexisting state occurs near $0.75\pi$ between FM and H-Zigzag states.

With increasing $\alpha$, the region of the H-N\'eel state decreases, while that of the nematic state increases, as evidenced from $O_\mathrm{N\acute{e}el}$ and $O_\mathrm{nematic}$ in Fig.~\ref{fig:energy}. There is no finite order parameters between the H-N\'eel and nematic states. The boundaries where the H-N\'eel and nematic order parameters disappear are hard to be determined from $-d^2E_\mathrm{unit}/d\theta^2$.

The coexisting state of H-N\'eel and stripy states in Fig.~\ref{fig:energy}(a) clearly persists with increasing $\alpha$ up to $\alpha=0.5$ near $\theta=-0.14\pi$. With further increasing $\alpha$, the coexistence disappears, but the reduction of $O_\mathrm{stripy}$ from 1 is clearly seen at $\theta=- \tan ^{-1} (1/2)\sim-0.147\pi$ even for $\alpha=1$. The reduction indicates the existence of  a coexisting state even for $\alpha=1$. This will be discussed in more detail below.

\begin{figure}[tb]
  \begin{center}
    \epsfig{file=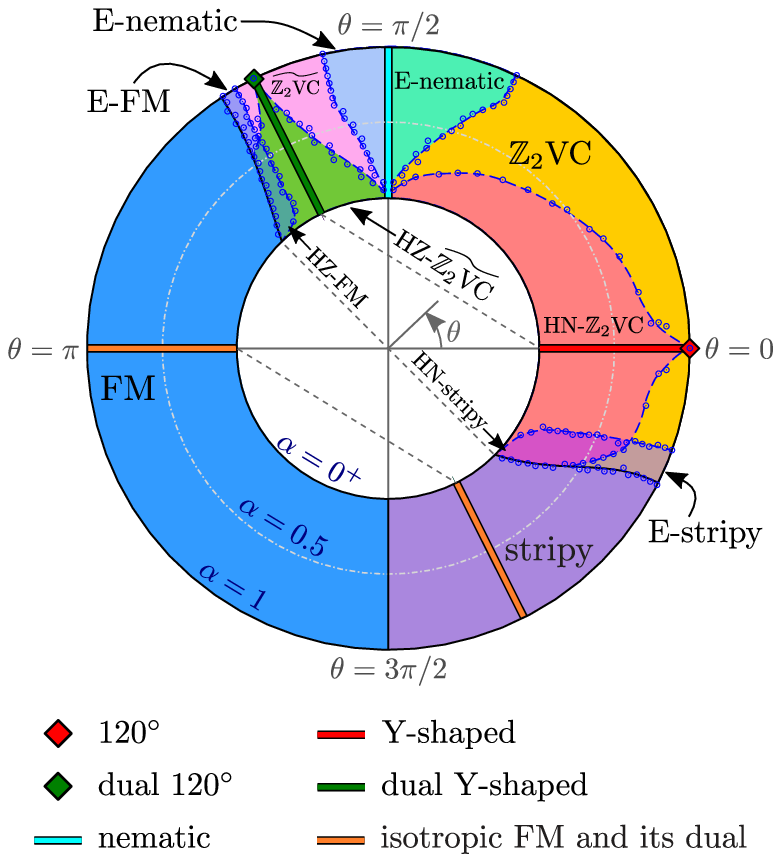, width=86mm}
    \caption{Classical ground state phase diagram of the KH model parametrized by $\alpha$ and $\theta$. The inner (outer) circle corresponds to $\alpha=0^+$ ($\alpha=1$) describing the honeycomb (triangular) lattice. There are various phases including their dual phases (see the text). The boundaries denoted by the black solid lines are determined by the LT method. The blue dots denote boundaries determined from order parameters calculated by the classical MC simulation for a $72\times 72 \times 3$-site lattice with PBC. The blue dashed lines interpolate the blue dots. Two $\theta$ values  inside circle connected by the dotted lines give the same state due to the Klein duality.}
    \label{fig:pdc}
  \end{center}
\end{figure}

We show the resulting classical ground state phase diagram in Fig.~\ref{fig:pdc}. From the inner circle to the outer circle, $\alpha$ changes from $\alpha=0^+$ (honeycomb lattice) to $\alpha=1$ (triangular lattice)  and thus increasing radius corresponds to increasing $\alpha$. At $\theta=0$, a Y-shaped spin state (see below) appears and changes to the so-called $120^{\circ}$ state at $\alpha=1$. Because of the Klein duality~\cite{kimchi2014},  dual Y-shaped spin state and dual $120^\circ$ state emerge at $\theta=\pi - \tan ^{-1}2$. At $\theta=\pi$, an isotropic FM state appears. At $\theta=\pi/2$, a nematic state appears for $0<\alpha\le 1$. In addition to these special cases, we identify other phases in Fig.~\ref{fig:pdc}: $\mathbb{Z}_2$ vortex crystal phase ($\mathbb{Z}_2$VC), its dual phase ($\widetilde{\mathbb{Z}_2\mathrm{VC}}$), extended-nematic phase (E-nematic), FM phase and its dual phase called stripy phase, extended-stripy (E-stripy) phase, and coexisting phases between H-N\'eel and $\mathbb{Z}_2$VC (HN-$\mathbb{Z}_2$VC), between H-zigzag and $\widetilde{\mathbb{Z}_2\mathrm{VC}}$ (HZ-$\widetilde{\mathbb{Z}_2\mathrm{VC}}$), between H-zigzag and FM (HZ-FM), and between H-N\'eel and strypy (HN-stripy). We will discuss characteristic behaviors on several phases.

\subsection{Y-shaped spin state and $120 ^{\circ}$ state}
\label{Y-shape}
\begin{figure}[tb]
  \begin{center}
	\epsfig{file=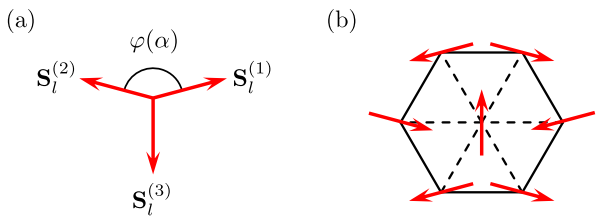, width=80mm}
    \caption{(a) Y-shaped spin state. The angle $\varphi$ is given by Eq.~(\ref{eq:y-shaped_angle}). (b) Snap shot of the ground state stabilized at $J=1$ and $K=0$. The ratio of interactions between the solid and dashed bonds is $1: \alpha$.}
    \label{fig:y-shaped}
  \end{center}
\end{figure}

At $\theta=0$, i.e., $(J,K)=(1,0)$, the model becomes the antiferromagnetic Heisenberg model. The three spins in the unit cell form a Y shape as shown in Fig.~\ref{fig:y-shaped}(a), where the angle between ${\bf S}_l ^{(1)}$ and ${\bf S}_l ^{(2)}$ is given by
\begin{equation}
  \varphi ({\alpha}) = \cos ^{-1} \left( -1 + \frac{\alpha ^2}{2} \right). \label{eq:y-shaped_angle}
\end{equation}
The $\varphi$ becomes $\pi$ for the honeycomb limit ($\alpha=0^+$) and $2\pi/3=120^\circ$ for the triangular lattice ($\alpha =1$). The ground state becomes a coplanar state with three sublattices as shown in Fig.~\ref{fig:y-shaped}(b), where $E_{\rm unit}=-3 \left( 1+ \frac{\alpha ^2}{2} \right)I_0$.

The ground state at $\theta=0$ has degeneracy characterized by the order parameter space ${\rm SO(3)}$ due to geometrical frustration as is the case of triangular lattice, and thus the first homotopy group is given by $\pi _1({\rm SO(3)})= \mathbb{Z}_2$~\cite{mermin1979,kawamura1984}. Therefore, similar to the previous study on the triangular lattice~\cite{becker2015,catuneanu2015,rousochatzakis2016}, a $\mathbb{Z}_2$ vortex crystal phase is expected in our  model when the Kitaev interaction is finite ($K \neq 0$).

\begin{figure*}[tb]
  \begin{center}
    \epsfig{file=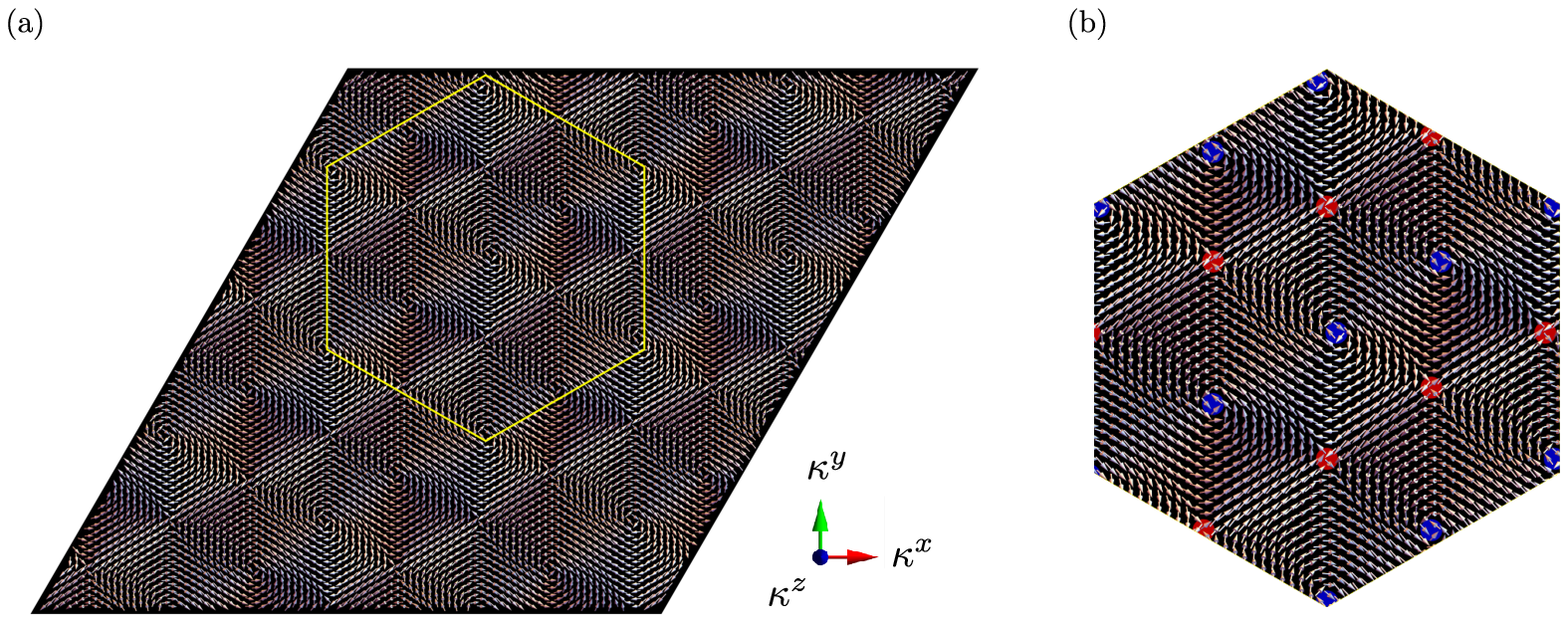, width=180mm}
    \caption{(a) Snap shot of vector chirality ${\bm \kappa}_i$ in the $\mathbb{Z}_2$VC phase of a $48 \times 48 \times 3$-site lattice with PBC obtained by the classical MC simulation. $\alpha=0.96$ and $\theta=-0.08$. An arrow on each site represents ${\bm \kappa}_i$ projected onto the $\kappa^x$-$\kappa^y$ plane. (b) Enlarged view of ${\bm \kappa}_i$ in the area surrounded by yellow lines in (a). The blue (read) dots denote the center of vortices (antivortices), which forms a crystal.}
    \label{fig:z2vc}
  \end{center}
\end{figure*}

For frustrated spin systems such as triangular Heisenberg model ($\alpha=1$ and $\theta=0$), vector chirality is a good quantity to characterize frustration~\cite{kawamura1984}. We define the vector chirality at site $i$ as
\begin{eqnarray}
{\bm \kappa}_i & \equiv & \frac{2}{\sqrt{(2- \alpha )(\alpha + 2)^3}} \nonumber \\
& & \ \ \ \times \left(\mathbf{S}_i \times \mathbf{S}_j + \mathbf{S}_j \times \mathbf{S}_k + \mathbf{S}_k \times \mathbf{S}_i \right),
\label{eq:chi}
\end{eqnarray}
where the site $j$ ($k$) is the nearest-neighbor site of $i$ with the displacement vector $(\mathbf{a}_1+\mathbf{a}_2)/3$ [$(-\mathbf{a}_1+2\mathbf{a}_2)/3$] from $i$ and thus the three sites, $i$, $j$, and $k$ form a triangle. The prefactor in Eq.~(\ref{eq:chi}) is chosen for the Y-shaped spin state to be $|{\bm \kappa}_i|=1$. We note that ${\bm \kappa}_i$ is aligned in the same direction independent of site $i$.

The Klein duality gives dual $120 ^{\circ}$ and dual Y-shaped states as shown in Fig.~\ref{fig:pdc}.

\subsection{$\mathbb{Z}_2$VC phase}
\label{Z2VC}

Away from $\theta=0$ and close to $\alpha=1$, there is the $\mathbb{Z}_2$VC phase that spreads in the region of $-0.1\pi\lesssim\theta\lesssim 0.4\pi$. In order to demonstrate  the presence of vortices, we calculate the vector chirality ${\bm \kappa}_i$ at $\alpha=0.96$ and $\theta=-0.08 \pi$, which is shown in Fig.~\ref{fig:z2vc} for a $N=48 \times 48 \times 3$-site lattice with PBC. Vortices and antivortices appear in pairs: the vortices are centered at the blue dots while the antivortecies are at the red dots in Fig.~\ref{fig:z2vc}(b). Since there is translational invariance, the vortex-antivertex pairs form a crystal. The averaged magnitude of the vector chirality given by $N^{-1}\sum_i |{\bm \kappa}_i|$ is $0.99$, which is close to 1, implying that spins locally tend to form the Y-shaped ordering. The main reason that the average value deviates from the ideal value 1 is the distortion of spin configuration from the Y shape around the cores. Thus, the cores of the $\mathbb{Z}_2$ vortices appear as the defects of the Y-shaped spin state. Similar to the case of triangular lattice~\cite{rousochatzakis2016}, the distance between the vortex cores becomes shorter as the $|K|$ becomes larger.

The Klein duality gives dual $\mathbb{Z}_2$VC phase that is denoted as $\widetilde{\mathbb{Z}_2\mathrm{VC}}$ in Fig.~\ref{fig:pdc}.

\subsection{HN-$\mathbb{Z}_2$VC phase}
\label{HN-Z2VC}

With decreasing $\alpha$ in the $\mathbb{Z}_2$VC phase, the H-N\'eel order emerges at a critical value of $\alpha$. In spite of the presence of the H-N\'eel order, the vortex-antivortex pairs remain. Therefore, we call the coexisting phase the HN-$\mathbb{Z}_2$VC phase.

\begin{figure}[tb]
  \begin{center}
    \epsfig{file=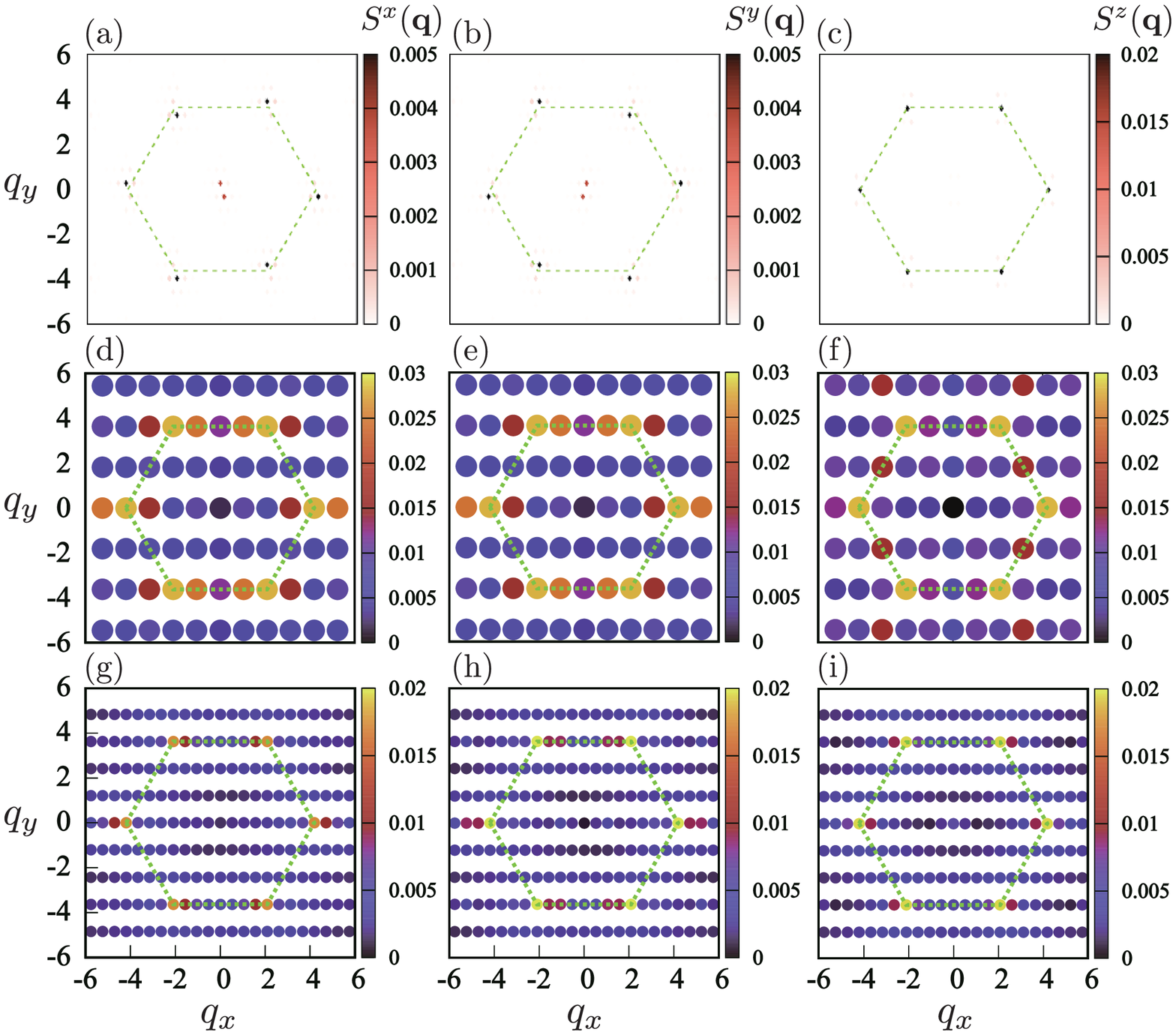, width=86mm}
    \caption{Spin structure factor in the HN-$\mathbb{Z}_2$VC phase in the extended Brillouin zone. $\theta=0.2\pi$ and $\alpha=0.2$. (a) $S^x(\mathbf{q})$, (b) $S^y(\mathbf{q})$, and (c) $S^z(\mathbf{q})$ in the classical system with a $48\times 48\times 3$-site lattice, obtained by the classical MC simulation. (d) $S^x(\mathbf{q})$, (e) $S^y(\mathbf{q})$, and (f) $S^z(\mathbf{q})$ in the quantum system with a 24-site lattice, obtained by the ED method. The circles denote the momentum positions defined in the lattice. (g) $S^x(\mathbf{q})$, (h) $S^y(\mathbf{q})$, and (i) $S^z(\mathbf{q})$ in the quantum system with a $12\times 6$-site lattice, obtained by the DMRG method. The circles denote the momentum positions defined by assuming PBC for the lattice. The green dotted hexagon in each panel represents the first Brillouin zone of the triangular lattice.}
    \label{fig:SqHNZ2VC}
  \end{center}
\end{figure}

\begin{figure}[tb]
  \begin{center}
    \epsfig{file=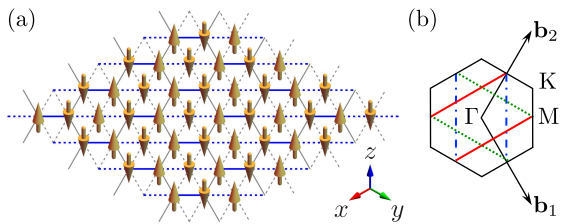, width=75mm}
    \caption{(a) One of spin configurations at the $3\times 2^L$-fold degenerated nematic phase at $\theta=\pi/2$, obtained by the LT method for a $4\times 4\times 3$-site lattice. The parameters in Eq.~(\ref{eq:nematic}) are set to $f_{n_1+2n_2+m}^x = 0$,
$f_{2n_1+n_2+m}^y = 0$, $f_0^z=-1$, $f_1^z=1$, $f_2^z=1$, and $f_3^z=-1$. (b) Wavenumbers ${\bf q}'$ that give the degenerate ground states in the nematic phase in the first Brillouin zone. The red solid, green dotted, and blue dashed-dotted lines represent $({\bf q}',x)$, $({\bf q}',y)$, and $({\bf q}',z)$, respectively. ${\bf b}_1$ and ${\bf b}_2$ are two primitive translation vectors of the reciprocal lattice. }
    \label{fig:nematic}
  \end{center}
\end{figure}

\begin{figure}[tb]
  \begin{center}
    \epsfig{file=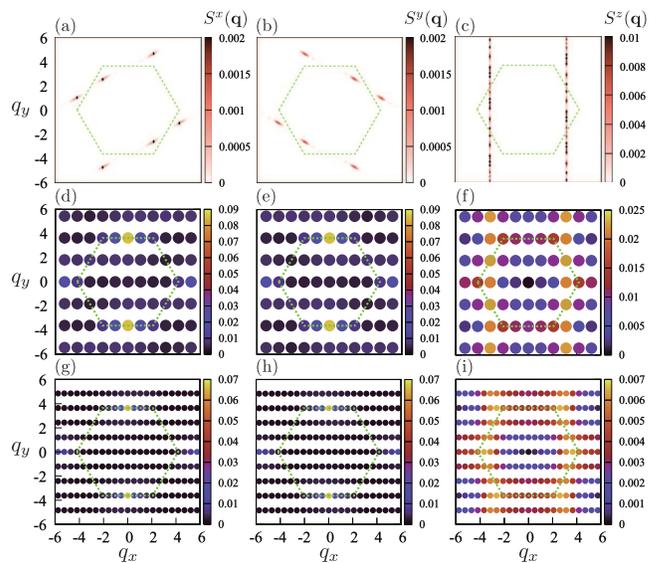, width=86mm}
    \caption{The same as Fig.~\ref{fig:SqHNZ2VC}, but in the E-nematic phase. $\theta=0.45\pi$ and $\alpha=0.8$.}
    \label{fig:SqEnematic}
  \end{center}
\end{figure}

\begin{figure}[tb]
  \begin{center}
    \epsfig{file=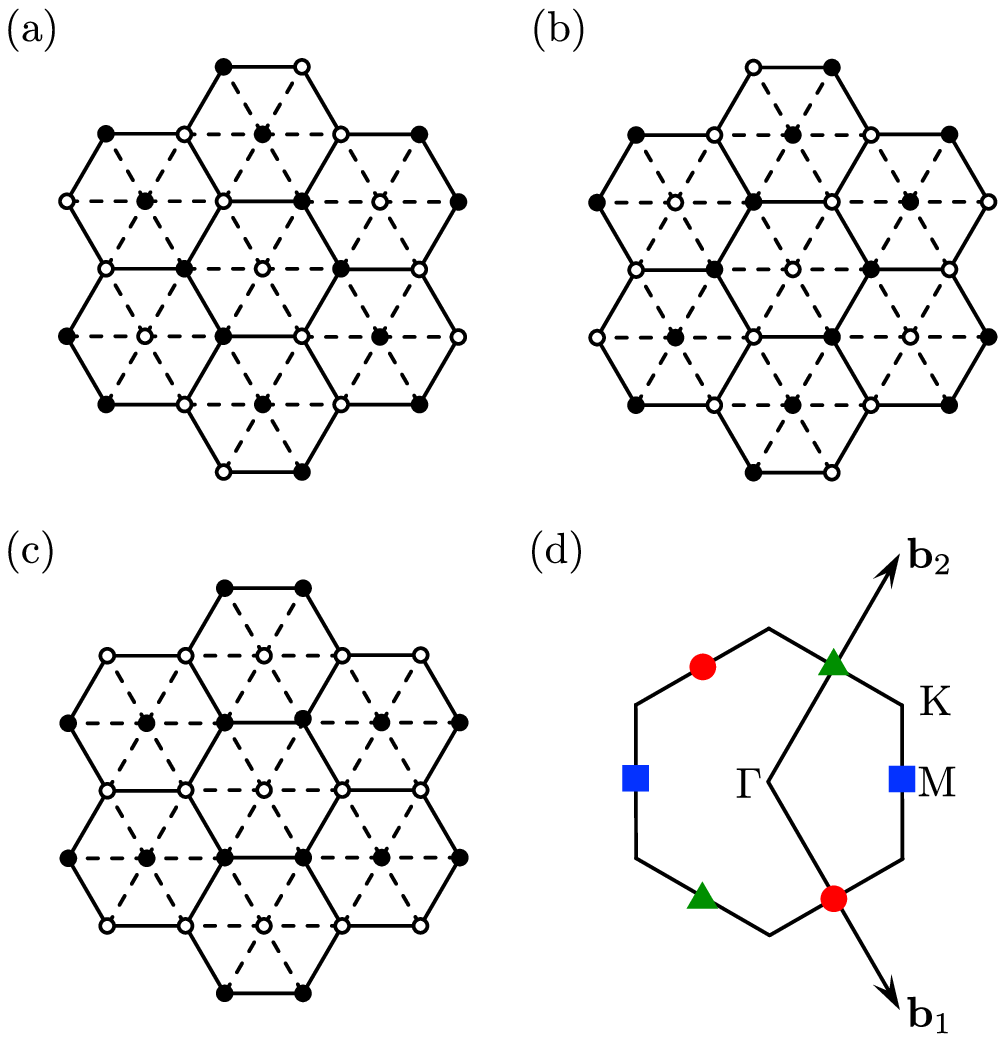, width=75mm}
    \caption{Schematic view of spin arrangement in the stripy phase obtained by the LT method. (a) $x$-component of spin $S_i ^x$, (b) $y$-component of spin $S_i ^y$, and (c) $z$-component of spin $S_i ^z$. The black and open circles represent opposite sign of $S_i ^\gamma.$ each other. The ratio of interactions between the solid and dashed bonds is $1: \alpha$. (d) Wavenumbers ${\bf q}'$ giving the ground state in the first Brillouin zone. The red circles, green triangles, and blue squares represent $({\bf q}',x)$, $({\bf q}',y)$, and $({\bf q}',z)$, respectively. All of these ${\bf q}'$ are on the M points. ${\bf b}_1$ and ${\bf b}_2$ are two primitive translation vectors of the reciprocal lattice. }
    \label{fig:dual-fm}
  \end{center}
\end{figure}

To confirm the coexistence of the H-N\'eel order and the $\mathbb{Z}_2$VC, we show $\mathbf{S}(\mathbf{q})$ obtained by the classical MC simulation for a $48\times 48\times 3$-site periodic lattice with $\theta=0.2\pi$ and $\alpha=0.2$ in Figs.~\ref{fig:SqHNZ2VC}(a)-\ref{fig:SqHNZ2VC}(c). $S^z(\mathbf{q})$ in Fig.~\ref{fig:SqHNZ2VC}(c) clearly shows strong intensity at the corner of the Brillouin zone, being consistent with the H-N\'eel state. $S^x(\mathbf{q})$ in Fig.~\ref{fig:SqHNZ2VC}(a) and $S^y(\mathbf{q})$ in Fig.~\ref{fig:SqHNZ2VC}(b) exhibit strong intensity slightly away from the corner, which is consistent with the signature of $\mathbb{Z}_2$VC discussed in the triangular lattice~\cite{becker2015}. In addition, we find two incommensurate structures around $\mathbf{q}=(0,0)$ in both $S^x(\mathbf{q})$ and $S^y(\mathbf{q})$. This is due to the remnants of FM order at $\theta=0$ seen in Figs.~\ref{fig:energy}(a) and \ref{fig:energy}(b), which are shifted to finite $\mathbf{q}$ by the presence of $\mathbb{Z}_2$VC.

As $\alpha$ decreases, the distance between vortex and antivortex increases accompanied with the reduction of the magnitude of ${\bm \kappa}_i$. At $\alpha=0^+$, the N\'eel phase on the honeycomb lattice appears~\cite{chaloupka2010,jiang2011,chaloupka2013}. 

The H-N\'eel state is changed to H-zigzag by the Klein duality. Therefore, the HN-$\mathbb{Z}_2$VC phase corresponds to the HZ-$\widetilde{\mathbb{Z}_2\mathrm{VC}}$ phase in Fig.~\ref{fig:pdc}.

\subsection{Nematic phase}
\label{nematic}
At $\theta= \pi/2$, i.e., $(J,K)=(0,1)$, the Hamiltonian (\ref{eq:hamiltonian_1}) becomes Kitaev model. In the triangular lattice ($\alpha=1$), a nematic state appears~\cite{becker2015,catuneanu2015,rousochatzakis2016}. Even for $0< \alpha <1$, we find the same nematic state by the LT method with $E_{\rm unit}=-(1+2 \alpha )I_0 $. In this phase, the spin configuration is given by
\begin{equation}
  {\bf S}_l ^{(m)}= \left(
  \begin{array}{rl}
    f_{n_1+2n_2+m} ^x & (-1)^{n_2+m} \\
    f_{2n_1+n_2+m} ^y & (-1)^{n_1+m} \\
    f_{-n_1+n_2} ^z & (-1)^{n_1+m}
  \end{array}
  \right), \label{eq:nematic}
\end{equation}
where the sets $\{ f_{n_1+2n_2+m} ^x \}$, $\{ f_{2n_1+n_2+m} ^y \}$, and $\{ f_{-n_1+n_2} ^z \}$ have the arbitrary numbers under the local constraints (\ref{eq:local}). Therefore, in classical level, the model has the subextensive ground state degeneracy as discussed by Rousochatzakis {\it et al.}~\cite{rousochatzakis2016}. Figure~\ref{fig:nematic} (a) shows one of spin configurations for a set of parameters listed in the caption of Fig.~\ref{fig:nematic}, where lines connected by blue solid and dotted lime segments denote the $\mu _z$ chains. We note that the $\gamma$  ($\gamma = x$, $y$, and $z$) component of spins in the nematic phase aligns antiferromagnetically along the $\mu _\gamma$ chain. Thus, each direction leads to $2^L$-fold states, where $L$ is the linear system size. The degenerate ground states are given by $\mathbf{q}'$ in the first Brillouin zone shown in Fig.~\ref{fig:nematic}(b). One of the three directions becomes lower in free energy, by, for example, thermal fluctuation, leading to the nematic state~\cite{rousochatzakis2016}.

\subsection{E-nematic phase}
\label{Enematic}

The nematic order parameter $O_{\rm n}$ in (\ref{eq:order_nematic}) calculated by the classical MC simulation is not unity but finite across $\theta = \pi /2$ for $0< \alpha \leq 1$ (see Fig.~\ref{fig:energy}). We call this phase the extended nematic (E-nematic) phase. This E-nematic phase narrows with the decrease of $\alpha$ and converges to a single point at $\alpha = 0^+$. 

$\mathbf{S}(\mathbf{q})$ is shown in Fig.~\ref{fig:SqEnematic} for $\theta=0.45\pi$ and $\alpha=0.8$. All of the $x$, $y$, and $z$ components have a two-line structure with nonuniform intensity, but the $z$ component is different form the $x$ and $y$ components. The two-line structure exhibits nematic behavior but not a perfect nematic state characterized by uniform lines as expected from Fig.~\ref{fig:nematic}(b). This is nothing but the remnants of the perfect nematic state.

\subsection{Stripy, E-stripy, and HN-stripy phases}
\label{stripy}
The stripy phase has been discussed in the honeycomb lattice~\cite{chaloupka2010,chaloupka2013,rau2014} as well as in the triangular lattice as dual FM phase~\cite{kimchi2014,becker2015,rousochatzakis2016,catuneanu2015}. The LT method gives the ground state with $E_\mathrm{unit}=-(1+2 \alpha )(J-K)$ at the M points shown in Fig.~\ref{fig:dual-fm}(d). These wavenumbers lead to the spin configuration given by
\begin{equation}
  {\bf S}_l ^{(m)}= \left(
  \begin{array}{rl}
    f^x & (-1)^{n_1+m} \\
    f^y & (-1)^{n_2+m} \\
    f^z & (-1)^{n_1+n_2}
  \end{array}
  \right), \label{eq:stripy}
\end{equation}
where $f^x$, $f^y$, and $f^z$ satisfy $(f^x)^2+(f^y)^2+(f^z)^2=1$ by the local constraints (\ref{eq:local}). Thus, in classical level, the order parameter space is $S^2$. The schematic view of spin arrangement is shown in Figs.~\ref{fig:dual-fm}(a), \ref{fig:dual-fm}(b), and \ref{fig:dual-fm}(c) for the $x$, $y$, and $z$ components of spin, respectively, where FM arrangement appears along the bond of the $\gamma$ components of the Kitaev interaction, forming a stripe shape of spin distribution. 

The stripy phase is the dual phase of FM phase through the Klein duality. Namely, $\theta = \tan ^{-1}2$ in the stripy phase corresponds to the pure FM Heisenberg model at $\theta = \pi$, which is represented by the orange line in Fig.~\ref{fig:pdc}. Therefore, the stripy state at $\theta = \tan ^{-1}2$ is the ground state even for quantum spins. When $\theta \neq \tan ^{-1}2$, there is the possibility that some degeneracy can be lifted by order-by-disorder mechanism because the corresponding Hamiltonian is not SU(2) symmetric. Indeed, it has been point outed that spins lie along one of the cubic axes on the triangular lattice~\cite{rousochatzakis2016, becker2015}.

The stripy order parameter $O_\mathrm{stripy}$ in (\ref{eq:order_stripy}) changes from one to smaller value, when $\theta$ approached to zero [see Figs.~\ref{fig:energy}(c) and \ref{fig:energy}(d)], forming a coexisting phase with $\mathbb{Z}_2$VC. We call this coexisting phase the extended stripy (E-stripy) phase. The distance between vortex and antivortex becomes smaller with approaching from $\mathbb{Z}_2$VC to the stripy phase. Because of finite size of the vortexes, the stripy order parameter becomes finite before the vortex-antivortex pair diminishes. This might be the origin of the coexisting phase. The dual phase of E-stripy is E-FM.

$O_\mathrm{stripy}$ also coexists with $O_\mathrm{H-N\acute{e}el}$ as shown in Figs.~\ref{fig:energy}(a) and \ref{fig:energy}(b). Therefore, we call this region the HN-stripy phase, whose dual phase is HZ-FM as shown in Fig.~\ref{fig:pdc}.

\begin{figure}[t]
  \begin{center}
    \epsfig{file=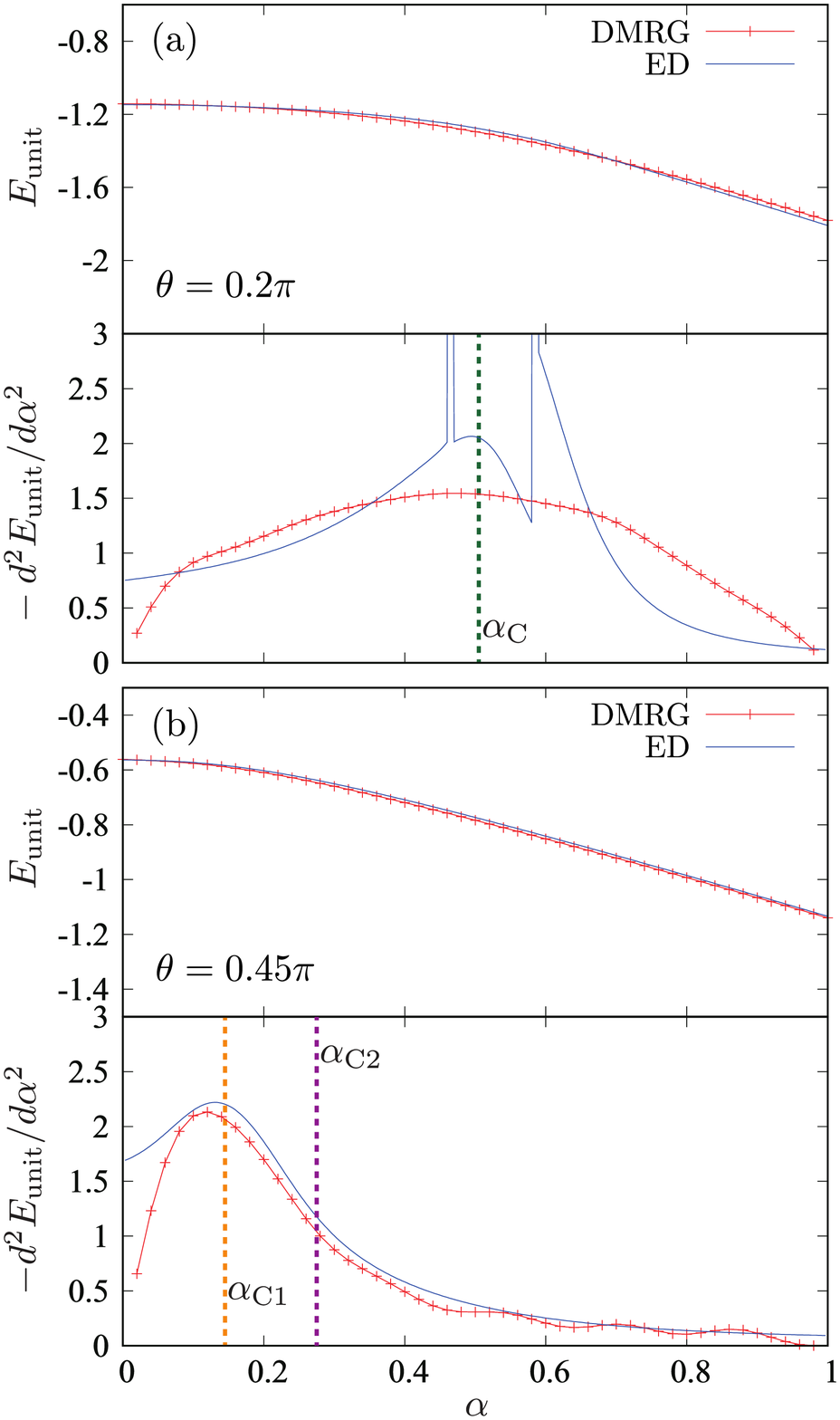, width=86mm}
    \caption{Quantum ground state energy per unit cell $E_{\rm unit}$ and second derivative of $E_{\rm unit}$ with respect to $\alpha$, $-d^2 E_{\rm unit}/d \alpha^2$, obtained by the ED method for a 24-site lattice and by DMRG for a $12\times 6$-site lattice.  (a) $\theta=0.2\pi$ and (b) $\theta=0.45\pi$. The vertical dotted lines correspond to phase boundaries obtained by the classical MC simulations.}
    \label{fig:qenergy}
  \end{center}
\end{figure}

\section{Quantum ground state phase diagram}
\label{Sec5}

In order to construct a quantum-mechanical phase diagram of the honeycomb-triangular KH model~(\ref{eq:hamiltonian_1}), we calculate the ground state energy by the ED method for the 24-site lattice. Figure~\ref{fig:qenergy} shows $E_\mathrm{unit}$, as a function of $\alpha$ and the second derivative with respect to $\alpha$, $-d^2E_\mathrm{unit}/d \alpha^2$ for given $\theta$. To check size dependence, we also show $E_\mathrm{unit}$ by DMRG for the $12\times 6$-site lattice with CBC in Fig.~\ref{fig:qenergy}. Both ED and DMRG results of $E_\mathrm{unit}$ show the $\alpha$ dependence similar to each other. The second derivative of $E_\mathrm{unit}$ at $\theta=0.2\pi$ shown in Fig.~\ref{fig:qenergy}(a) exhibits two anomalies around $\alpha=0.5$ in ED but does not in DMRG. The anomalies come from level crossing due to small system size. However, their positions as well as a broad maximum in the DMRG result are closely located at the classical phase boundary denoted by $\alpha_\mathrm{C}$ in Fig.~\ref{fig:qenergy}(a). Therefore, a phase boundary is expected to exist nearby the classical one. The results at $\theta=0.45\pi$ shown in Fig.~\ref{fig:qenergy}(b) also exhibit possible phase boundary close to the classical one denoted by $\alpha_\mathrm{C1}$, though anomaly around $\alpha_\mathrm{C2}$ is unclear.

\begin{figure}[tb]
  \begin{center}
	\epsfig{file=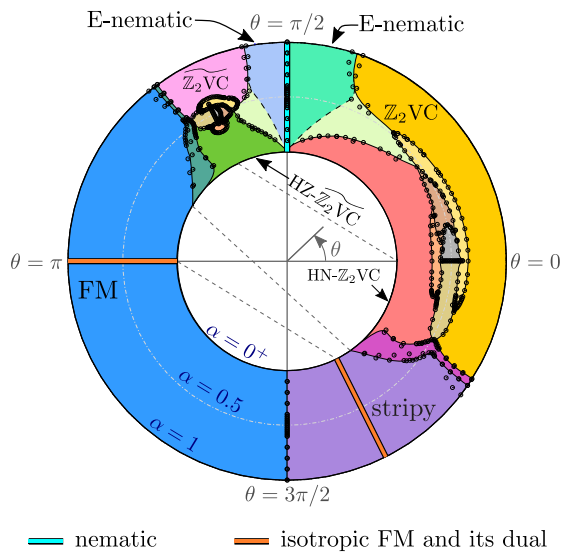, width=86mm}
    \caption{Quantum ground state phase diagram of the KH model parametrized by $\alpha$ and $\theta$, obtained by the ED method for a 24-site lattice with PBC. The inner (outer) circle corresponds to $\alpha=0^+$ ($\alpha=1$) describing the honeycomb (triangular) lattice. There are various phases including their dual phases similar to the classical phase diagram shown in Fig.~\ref{fig:pdc}. Two value of $\theta$ inside circle connected by the dotted lines give the same state as a result of the Klein duality.}
    \label{fig:pdq}
  \end{center}
\end{figure}

Taking into account the second derivative of $E_\mathrm{unit}$ as well as $\mathbf{S}(\mathbf{q})$ (not shown), we construct a phase diagram by the ED method as shown in Fig.~\ref{fig:pdq}. The phase boundaries are drown by interpolating some of points that indicate the change of phases. We can recognize similarity to the classical phase diagram in Fig.~\ref{fig:pdc} at a glance. 

However, there are several differences between the two phase diagrams. The most prominent difference is a complicated  distribution of many phases around the region of $\alpha\sim 0.5$ and $-0.2\pi\lesssim\theta\lesssim 0.25\pi$ together with its dual region. Along the $\theta=0$ line, the total spin of the system changes from zero at $\alpha=1$ to discreet finite values with decreasing $\alpha$, and then return to zero for $\alpha<0.32$. Those states with finite total spin at $\theta=0$ are smoothly connected to the phases showing the complicated distribution. Another difference between the classical and quantum phase diagrams appears at the E-nematic phase. In the classical case, the boundary separating the E-nematic and $\mathbb{Z}_2$VC phases reach at the Kitaev spin liquid point, i.e., $\theta=\pi/2$ and $\alpha=0$. In contrast, the boundary in the quantum case extends toward smaller $\theta$ starting from $\theta\sim0.4\pi$ at $\alpha=1$. However, we find a point along the boundary where the transition changes from a weakly first-order type to a second-order type. Therefore, we expect a possible point where the character of the boundary changes and thus we draw the broken line from the possible point to the Kitaev spin-liquid point. Of course, the phase diagram obtained by ED suffers from the finite-size effect. DMRG would be a possible alternative method, but unfortunately numerical costs to make a complete phase diagram are demanding. Making the complete phase diagram in the quantum system remains as a future problem.

We also examine $\mathbf{S}(\mathbf{q})$ by the ED and DMRG methods and compare them with the classical results. Figures~\ref{fig:SqHNZ2VC} and \ref{fig:SqEnematic} shows $\mathbf{S}(\mathbf{q})$ in the HN-$\mathbb{Z}_2$VC and E-nematic phases, respectively. The ED and DMRG results are similar to each other, indicating small system-size dependence. In the E-nematic phase, we find a $\mathbf{q}$ direction with large intensity in $S^z(\mathbf{q})$ [see Figs.~\ref{fig:SqEnematic}(f) and ~\ref{fig:SqEnematic}(i)], which is qualitatively consistent with the classical one in Fig.~\ref{fig:SqEnematic}(c). This means that both the quantum and classical systems have the same E-nematic phase. The ED results of $S^x(\mathbf{q})$ in Fig.~\ref{fig:SqEnematic}(d) and $S^y(\mathbf{q})$ in Fig.~\ref{fig:SqEnematic}(e) show the same behavior. The DRMG results in Figs.~\ref{fig:SqEnematic}(g) and \ref{fig:SqEnematic}(h) are also similar. These similarities, however, are not seen in the classical results in Figs.~\ref{fig:SqEnematic}(a) and \ref{fig:SqEnematic}(b). The difference indicates that the degeneracy between $S^x(\mathbf{q})$ and $S^y(\mathbf{q})$ is not lifted in quantum systems. In the HN-$\mathbb{Z}_2$VC phase, the HN feature is clearly seen as evidenced by strong intensity at the corner of the Brillouin zone in $S^z(\mathbf{q})$ [see Figs.~\ref{fig:SqHNZ2VC}(f) and \ref{fig:SqHNZ2VC}(i)], being consistent with the classical case in Fig.~\ref{fig:SqHNZ2VC}(c). However, the evidence of $\mathbb{Z}_2$VC is unclear because the intensity of $S^x(\mathbf{q})$ in Figs.~\ref{fig:SqHNZ2VC}(d) and \ref{fig:SqHNZ2VC}(g) as well as that of $S^y(\mathbf{q})$ in Figs.~\ref{fig:SqHNZ2VC}(e) and \ref{fig:SqHNZ2VC}(h) is distributed differently from the classical ones. This is probably due to small system size along the $y$ direction in quantum cases.

\section{Summary}
\label{Sec6}
In summary, we constructed a model connecting the honeycomb and triangular KH lattices and examined the ground state of the classical system by the LT method and the classical MC simulation. We found coexisting phases in the honeycomb-triangular lattice, which are composed of known phases in the honeycomb and triangular lattices. The quantum effects on the honeycomb-triangular lattice were examined by the ED and DMRG methods. The phase boundaries in the classical phase were suggested to survive even though quantum fluctuations are introduced as demonstrated by the phase diagram obtained by the ED method. However, the quantum phase diagram suffers from the finite-size effect. Therefore, it is desired to perform a systematic study increasing the system size, which is remains as a future problem.

\begin{acknowledgments}
We thank K. Shinjo, Y. Yamaji, T. Okubo, N. Kawashima, and M. Imada for useful discussions. 
This work was supported by MEXT, Japan, as a social and scientific priority issue (creation of new functional devices and high-performance materials to support next-generation industries) to be tackled by using a post-K computer and by MEXT HPCI Strategic Programs for Innovative Research (SPIRE) (hp140136, hp150142, hp160122).  This work has also been
supported by Grant-in-Aid for Scientic Research from MEXT Japan under the Grant No. 17K14148. The numerical calculation was carried out at the K Computer and HOKUSAI, RIKEN Center for Computational Science, and the facilities of the Supercomputer Center, the Institute for Solid State Physics, the University of Tokyo.
\end{acknowledgments}

\end{document}